\documentclass[aps,prl,preprint,groupedaddress,showpacs]{revtex4}

\usepackage{graphicx}
\usepackage{units}
\usepackage{cancel}

\newcommand{\bea}{\begin{eqnarray}}
\newcommand{\eea}{\end{eqnarray}}
\newcommand{\be}{\begin{equation}}
\newcommand{\ee}{\end{equation}}

\begin{document}

\title{Neutron Charge Density from Simple Pion Cloud Models}

\author{Jared A. Rinehimer}
\author{Gerald A. Miller}
\affiliation{Department of Physics, University of Washington, Seattle, Washington 98195-1560, USA}

\date{\today}

\begin{abstract} The physical nucleon is modeled as a bare nucleon surrounded by a pion cloud using
a psuedoscalar pion-nucleon coupling  to
examine its implications for the neutron form factor, $F_{1}(Q^{2})$, and
the corresponding transverse charge density, $\rho(b)$ in the infinite
momentum frame. Two versions, one with a tunable Pauli-Villars parameter,
and one with light-front cloudy-bag pion-nucleon form factors, are
examined. A qualitative agreement with the experimental form factor
($F_1<0$ for all $Q^2$) and transverse charge density are achieved
when the nucleon is treated as having a finite extent. The bare
nucleon must have a finite extent if this model is to account for a
negative definite $F_1$, and consequently a negative transverse charge
density of the neutron at its center.    
\end{abstract}


\maketitle

\section{Introduction}
It has been recently demonstrated that the neutron transverse charge
density is negative at the center \cite{MillerPrl07}. This surprising
result  seems to contrast with  
many long-standing arguments such as the 1947
predictions provided Fermi and Marshall, meson cloud models, and
arguments from gluon exchange \cite{Fermi1947, Thomas1981, Friar1972,
  Carlitz1977, Isgur1981}. As such, this
discovery requires new attempts at understanding the physics behind
the currently observed transverse density. This paper seeks to use
specific toy models to predict a qualitatively similar structure in
order to shed light on the physics involved.

The nucleonic form factors are defined by the matrix element of the
electromagnetic current operator $J^{\mu}(x)$ as
\be
\label{ff}{\langle}p',{\lambda}'|J^{\mu}(0)|p,{\lambda}{\rangle}= \nonumber
\bar{u}(p',{\lambda}')[{\gamma}^{\mu}F_1(Q^2)+i{\frac{{\sigma}^{{\mu}{\nu}}q_{\nu}}{2M}}F_2(Q^2)]u(p,\lambda) , 
\ee 
where $q=p'-p$ is the momentum transfer which is space-like such that
$0<-q^2 \ {\equiv} \ Q^2$. These form factors have the common
interpretation of being related to the charge and magnetization
density of the nucleon in question, however such an interpretation is
only valid at zero momentum transfer for which the initial and final nucleon wave functions are the same. 
In the infinite momentum frame (IMF), however, the
transverse charge density has been shown to be the two dimensional
Fourier transform of the form factor $F_1(Q^2)$
\cite{Soper:1976jc,Burkardt:2002hr,Diehl:2002he,MillerPrl07,Carlson:2007xd}. Production of these form factors in a
model will have implications for the physics involved with the
negative transverse charge density of the neutron.    

\section{Models}
\begin{figure}
\includegraphics{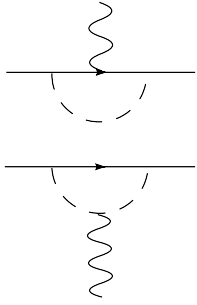}
\caption{\label{feynman} The two diagrams for the neutron-photon
  interaction to one loop order. The dashed lines represent pion
  propagators, the solid represent nucleon propagators.}\end{figure}

In this paper, we model the physical nucleon as a bare nucleon
surrounded by a pion cloud using psuedoscalar pion-nucleon coupling
\cite{Bjorken},  starting with the bare nucleon assumed to be a  point particle. To one-loop order, two diagrams exist for
the interaction of the neutron with a virtual photon, as shown in
Fig. \ref{feynman}.
\subsection{Model 1}
\begin{figure}
\includegraphics{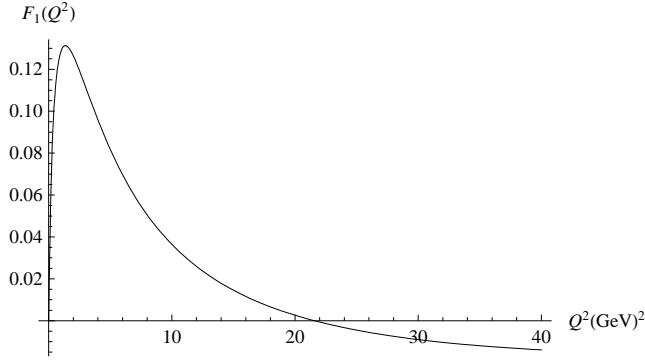}
\caption{\label{F1.1} The form factor $F_1(Q^2)$ for the Pauli-Villars
  model, with $\Lambda=1.0$ GeV. It is important to note that $F_1>0$ for some values of $Q^2$, while the observed $F_1(Q^2)$ is negative for all $Q^2$. }
\end{figure}

For the  first model, we introduce a Pauli-Villars regulator on the pion,
with the result that in the calculation of each diagram, one subtracts
a term identical to the diagram with the exception of the pion mass
being replaced by the regulator mass $\Lambda$. This regulator is then
used as a tunable parameter for the theory, with the advantage that this procedure
is co-variant and does not alter the Ward identities. The first
diagram in Fig. \ref{feynman}, evaluates to 
\bea
\left(4e{g_0}^2\bar{u}(p')\int\,dx\,dy\,dz\frac{d^4\ell}{(2\pi)^4}\delta(1-x-y-z)\left[\frac{{\nicefrac{\ell^2}{2}}-(x+y)^2M^2+xyQ^2)}{(\ell^2-\Delta_{m_\pi})^3}\gamma^\mu\nonumber\right.\right. \\ 
\left.\left.+\frac{2M^2(x+y)^2}{(\ell^2-\Delta_{m_\pi})^3}\frac{i\sigma^{\mu\nu}}{2M}q_\nu)\right]u(p)\right)\,-\,(m_\pi\leftrightarrow\Lambda) \, \label{feyn1}
\eea
where
\be
\Delta_{m_\pi} = z\,{m_\pi}^2+(x+y)^2M^2+xyQ^2,
\ee
and $\epsilon_\mu^*$ is the polarization of the virtual photon and is
omitted for the calculation of the matrix elements of  $J^\mu(0)$, $p$
is the incoming 4-momentum, $p'$ is the outgoing 4-momentum, $M$ is
the neutron mass and $m_\pi$ is the pion mass($\pi^-$).The second diagram is
\begin{eqnarray}
\left(4e{g_0}^2\bar{u}(p')\int\,dx\,dy\,dz\frac{d^4\ell}{(2\pi)^4}\delta(1-x-y-z)\left[\frac{{\nicefrac{-\ell^2}{2}}-2M^2z^2}{(\ell^2-\Delta'_{m_\pi})^3}\gamma^\mu \right.\right.\nonumber\\\left.\left.+\frac{2M^2z^2}{(\ell^2-\Delta'_{m_\pi})^3}\frac{i\sigma^{\mu\nu}}{2M}q_\nu)\right]u(p)\right)\,-\,(m_\pi\leftrightarrow\Lambda) \,\label{feyn2}
\end{eqnarray}
where
\begin{equation}
\Delta'_{m_\pi} = M^2z^2+(x+y){m_\pi}^2+xyQ^2.
\end{equation}
By examining (\ref{ff}), the terms proportional to $\gamma^\mu$
correspond to $F_1$, while the terms proportional to
$\frac{i\sigma^{\mu\nu}}{2M}q_\nu$ correspond to $F_2$. We can
evaluate the integral over the loop momentum $\ell$ by Wick rotation,
and the divergent terms cancel due to the Pauli-Villars
procedure. Then, we must sum both diagrams together for the full
contribution, giving expressions for the form factors
\begin{eqnarray}
F_1(Q^2)=\frac{{g_0}^2}{8\pi^2}\int_0^1\int_0^{1-x}\,dx\,dy \left[\text{ln}\left(\frac{\Delta_\Lambda\Delta'_{m_\pi}}{\Delta_{m_\pi}\Delta'_\Lambda}\right)\right.\nonumber\\
+\left((x+y)^2M^2-xyQ^2\right)\left(\frac{1}{\Delta_{m_\pi}}-\frac{1}{\Delta_\Lambda}\right)\\
\left.+2M^2(1-x-y)^2\left(\frac{1}{\Delta'_{m_\pi}}-\frac{1}{\Delta'_\Lambda} \right)\right],\nonumber
\end{eqnarray}
and,
\begin{eqnarray}
F_2(Q^2)=-M^2\frac{{g_0}^2}{4\pi^2}\int_0^1\int_0^{1-x}\,dx\,dy\left[(x+y)^2\left(\frac{1}{\Delta_{m_\pi}}-\frac{1}{\Delta_\Lambda}\right)\right. + \left.(1-x-y)^2 \left(\frac{1}{\Delta'_{m_\pi}}-\frac{1}{\Delta'_\Lambda}\right)\right]
\end{eqnarray} 
Numerical integration over Feynman parameters yields a functional form
for $F_1(Q^2)$ and $F_2(Q^2)$, which is used to derive a numerical
interpolating function. We use a neutron mass of $M=0.93957$ GeV and
pion mass of $m_{\pi}=0.13957$ GeV with a coupling of
$\frac{{g_0}^2}{4\pi}=13.5$. The value of $\Lambda=1.0$ GeV is
selected since it yields $F_2(0)=-1.9091$, close to the expected
magnetic moment of the neutron. Also numerical integration verifies
that $F_1(0)=0$, which agrees with charge conservation. A plot of
$F_1(Q^2)$ appears in Fig \ref{F1.1}.
\begin{figure}
\includegraphics{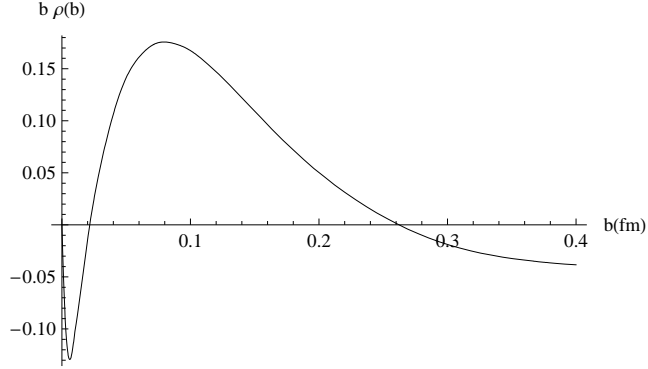}
\caption{\label{chg} The transverse charge density in the infinite
  momentum frame, derived from the form factor from Fig. \ref{F1.1}.}
\end{figure}

As stated before, the two-dimensional Fourier transform of $F_1$ is
the charge density in the IMF. Since $F_1$
lacks angular dependence, this density becomes 
\begin{equation}
  {\rho}(b)={\int}\frac{Q\,dQ}{2\pi}F_1(Q^2)J_0(bQ). \label{rhob}\end{equation} 
where $J_0$ is a Bessel function of the first kind, $b$ is the
transverse distance from the origin. Using the
interpolation function $F_1(Q^2)$, we find the charge density by
integrating to a sufficiently high cutoff. A plot of this charge density
appears in Fig. \ref{chg}. The density  is
negative at the center $(b=0)$, becomes positive as the value of $b$ increases, and then at still larger values of 
$b$ has a long, negative tail as found in ~\cite{MillerPrl07}. This
indicates in this model that the pion wave-function possesses some
non-zero probability to exist at the center of the neutron, since the
only negatively charged constituent of each Feynman diagram is the
$\pi^-$. This model however, yields a form factor for the neutron that
disagrees with observed data. In particular, $F_1>0$ for some values of $Q^2$,
while the observed $F_1$ is negative definite.

\subsection{Model 2} 
\begin{figure}
\includegraphics{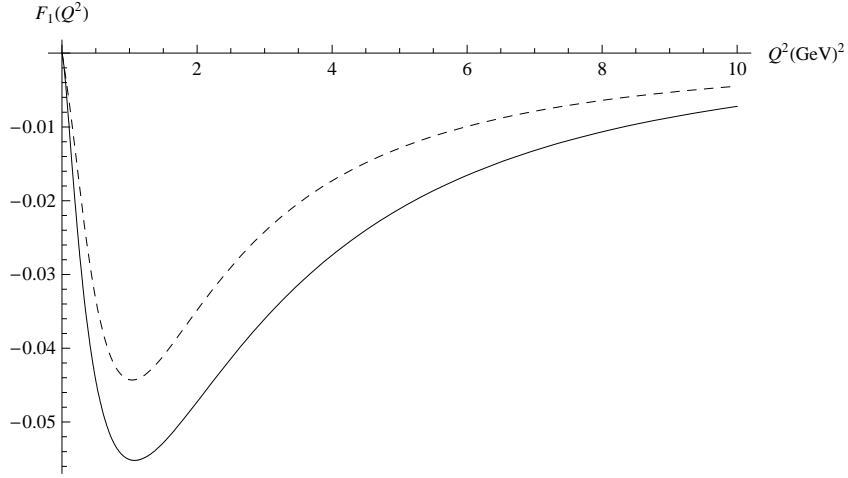}
\caption{Solid: The form factor $F_{1}$ for the neutron, using the value
  $r_s^2=10 (\mbox{GeV})^{-2} \approx  0.39 \mbox{fm}^2$. Dashed: The
  parametrization of \cite{Kelly2004}.\label{F1rs}}
\end{figure}

\begin{figure}
\includegraphics{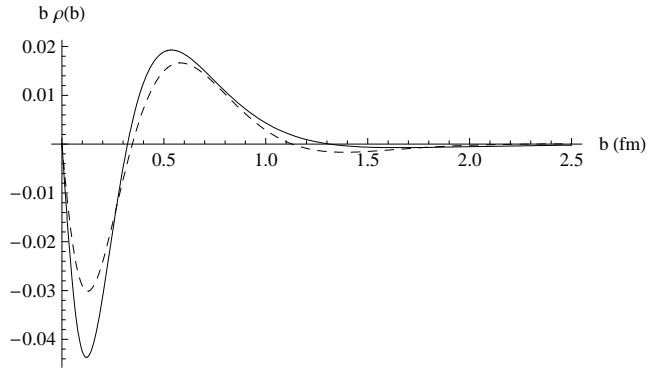}
\caption{Solid: The transverse charge density for the form factor from
  Fig. \ref{F1rs}. Dashed: The transverse charge density for the
  parametrization of \cite{Kelly2004}. \label{rhors}}
\end{figure}

Another procedure is to  
calculate the
diagrams by using a distinct form factor at each pion-nucleon
vertex. Then we use the light-front coordinates and an integration
over the $k^-$ variable to find numerically integrable expressions for
the form factor contributions from each Feynman diagram in
Fig. \ref{feynman}. The first diagram in Fig. \ref{feynman} gives a
term contributing to $F_1(Q^2)$ of the neutron as
\be
F_{n1,1}(Q^2)={g_0}^2\int_0^1\,dx\int \frac{d^2L}{(2\pi)^3}
R({\mathbf{L}_+}^2,x) R({\mathbf{L}_-}^2,x)(x^2 (M^2-Q^2/4)+\mathbf{L}^2)
\label{F11}\ee 
while the second diagram gives
\be
F_{n1,2}(Q^2)=-{g_0}^2\int_0^1\,dx\int
\frac{d^2K}{(2\pi)^3}R({\mathbf{K}_+}^2,x)R({\mathbf{K}_-}^2,x)(\mathbf{K}^2+M^2x^2-(1-x)^2Q^2/4)
\ee
where $\mathbf{L}$ is a shifted transverse momentum integration
parameter in the light-front coordinates, $\mathbf{L}_\pm = \mathbf{L}
\pm x\mathbf{q}/2$, $x=k^+/p^+$, $k^\mu$ is the virtual pion momentum,
$R(k^2,x)=F_{\pi N}(\mathbf{k}_\perp^2,x)/D(\mathbf{k}_\perp^2,x)$,
$D(\mathbf{k}_\perp^2,x)=M^2x^2+\mathbf{k}_\perp^2+m_\pi^2(1-x)$, and
the nucleon-pion form factor $F_{\pi N}$ is given by
\cite{ZollerZphys1992,HoltNuclA1996,Miller:2002ig,MatevPrc2005}
\be
F_{\pi N}(\mathbf{k}_\perp^2,x)=e^{-D(\mathbf{k}_\perp^2,x)/2x(1-x)\Lambda'}.
\ee
Finally, on the first term of $F_{1n}$ for the
neutron,(\ref{F11}), we multiply by a factor
\be
\chi=\frac{1}{(1+Q^2r_s^2/12)^2}, \label{ff_term1}
\ee
where $r_s$ is a free radius parameter. This factor corresponds to
granting an extent to the bare proton in the first diagram of
Fig. \ref{feynman}. Without this introduction, the form factor would be 
positive for some small values of $Q^2$, which does not agree with
observation. The expression for the form factors are then numerically
integrated over $\mathbf{L}^2$ and $x$ to give the behavior versus
$Q^2$ only. The final expression yields a form factor $F_{1n}$ for the neutron
that is qualitatively similar to that of existing parameterizations
\cite{Kelly2004,Bradford2006}. The form factor with $r_s^2=10.0\;
\mbox{GeV}^{-2}\approx 0.39 \;\mbox{fm}^2$,
$\Lambda=1.25\; \mbox{GeV}$, ${g_0}^2/{4\pi}=13.5$, and the
parameterization from \cite{Kelly2004} appear in Fig. \ref{F1rs}.
The  corresponding  transverse charge density, the two-dimensional Fourier transform of this version of $F_1$,
 is obtained from Eq.~(\ref{rhob}) , and appears in Fig. \ref{rhors}.
The transverse charge density from Model 2 qualitatively agrees with
the current observation that $\rho(b)$ is  negative at the center, is positive in the middle and negative at the far edge. 
\section{Summary}
The form
factor for Model 2 agrees with the observation that the measured form factor is negative for
all values of $Q^2$.  Achieving this occurs  only with the insertion of a term that accounts for 
the finite spatial extension of the bare nucleon.  Without the
introduction of the factor (\ref{ff_term1}), the $F_1$ function and the
transverse charge density $\rho(b)$ look qualitatively similar to that of
Fig. \ref{F1.1} and Fig. \ref{chg}, respectively.  Thus accounting for
the finite extension of the bare nucleon is necessary
to achieve a qualitatively reasonable treatment of the neutron's $F_1$ within the present  framework.

The present model is not realistic because the   factor (\ref{ff_term1}) is not computed from an underlying model and because proton properties are not studied.
Future work will be concerned with obtaining a better, more rigorous
formalism that outlines possible physical justifications for the observed
 transverse charge distribution within the neutron.
\section*{Acknowledgments}
We thank the USDOE for partial support of this work.
\bibliography{densityfin}
\end{document}